\documentclass[12pt]{iopart}

\usepackage{graphicx,epsfig}
\usepackage{times,amssymb,latexsym}
\usepackage{rotating}

\begin{document}

\title[Quantum simulation of  conductivity plateaux and fractional quantum Hall effect \dots]{Quantum simulation of  conductivity plateaux and fractional quantum Hall effect using ultracold atoms}

\author{Nuria Barber\'an$^1$, Daniel Dagnino$^2$, Miguel Angel Garc\'{\i}a-March$^{1,3}$, Andrea Trombettoni$^{4,5}$, Josep Taron$^{1,6}$, Maciej Lewenstein$^{3,7}$}

\address{$^1$Departament d'Estructura i Constituents de la Mat\`eria, Facultat de F\'isica, Universitat de Barcelona, Diagonal 645, E-08028 Barcelona, Spain}
\address{$^2$Barcelona Center for Subsurface Imaging, Institute of Marine Sciences, CSIC, Pg. Mar\'itim de la Barceloneta 37-49, 08003  Barcelona, Spain}
\address{$^3$ICFO-Institut de Ci\`encies Fot\`oniques, The Barcelona Institute of Science and Technology, 08860 Castelldefels (Barcelona), Spain}
\address{$^4$CNR-IOM DEMOCRITOS Simulation Center, Via Bonomea 265, I-34136 Trieste, Italy}
\address{$^5$SISSA and INFN, Sezione di Trieste, Via Bonomea 265, I-34136 Trieste, Italy}
\address{$^6$Institut de Ci\`encies del Cosmos, E-08028 Barcelona, Spain. }
\address{$^7$ICREA-Instituci\'o Catalana de Recerca i Estudis Avan\c{c}ats, 08010 Barcelona, Spain}

\begin{abstract}

We analyze the role of impurities in the fractional quantum Hall effect using a highly controllable system of ultracold atoms.  
We investigate the mechanism responsible for the formation of plateaux in the resistivity/conductivity as a function of the applied magnetic field  in the lowest Landau level regime.
To this aim, we consider an impurity immersed in a small cloud of an ultracold quantum Bose 
gas subjected to an artificial magnetic field. We consider scenarios corresponding to 
experimentally realistic systems with gauge fields induced by rotation of the trapping parabolic potential. Systems of this kind are adequate to simulate quantum Hall effects in ultracold atom setups. 
We use exact diagonalization for few atoms and, to emulate 
transport equations, we analyze the time evolution of the system under a periodic perturbation. We provide 
a theoretical proposal to detect the up-to-now elusive presence of strongly 
correlated states related to fractional filling factors in the context of ultracold atoms. 
We analyze the conditions under which these strongly correlated states are associated 
with the presence of the resistivity/conductivity plateaux. Our main result is the presence of a plateau 
in a region, where the transfer between localized and non-localized particles takes place, 
as a necessary condition to maintain a constant value of the resistivity/conductivity as the magnetic field increases. 

\end{abstract}

\date{\today }
\maketitle

\section{Introduction}

Transport properties provide some of the most fundamental characteristics of condensed matter systems 
(cf.~\cite{Imry1,Imry2}). 
In contrast, in physics of ultracold atomic and molecular gases \cite{Lew2007}, the studies of 
transport, unlike those performed in solid state settings, are hindered by the difficulty of having continuous and durable 
flow of atoms; for this reason they have been very limited so far. Among others they included: the investigations of Bloch oscillations 
(from the early studies with cold atoms~\cite{Ben1996} to the recent experiments with disordered gases \cite{Ree2014}), the extensive work on transport and diffusion in disordered gases~\cite{Luc2011,Der2013,Luc2013,Tan2013,Sem2014}, 
and the very recent experiments on quantized conductivity~\cite{Thy1999,Kri2013,Sta2012,Bra2012,Kri2014}. 

Paradigmatic systems, in which the transport properties play an essential role, are the systems that 
exhibit integer or fractional quantum Hall effects (IQHE, or FQHE) \cite{yos,Pra1987}. 
The quantum Hall effect consists in fact in {\it quantisation of the transverse conductance} 
for electronic current in the condensed matter systems, and for atomic flow for neutral atomic gases. 
Although in the IQHE the interactions play an irrelevant role, the underlying physics, 
even if well understood, is highly non-trivial. The case of the FQHE, where the interaction 
has a crucial contribution, is more complex and not yet completely understood. 
For these reasons systems that exhibit  FQHE belong to the most popular systems of strongly correlated particles that 
still await conclusive explanations and ``call for"  quantum simulations, for instance with ultracold atoms or 
ions~\cite{Lew2007}. 

In order to quantum simulate QHE it is necessary to generate strong artificial magnetic (gauge) fields. 
In the context of ultracold atomic systems, first realizations of artificial gauge (magnetic) 
fields were considered in rotating traps~\cite{Wil2000,Coo2001,Par2001,Bar2005a,Bar2006,Dag2009,Grass2011a}. 
Quite soon it was realized, however, that the most promising way to generate the artificial gauge fields 
is to use the laser induced  fields -- these methods are described in detail in several 
reviews~\cite{Lew2007,Blo2008,Dal2011,Gol2013}, while for the recent spectacular experiments 
the reader should consult cf.~\cite{Bee2013,Ken2013,Aid2013,Aid2014}. Various methods of detection of 
the Hall effect have been proposed and realized in the mentioned experiments 
(see cf.~\cite{Ber2010,Grass2012a}). In particular, 
it was shown~\cite{Uhu2008} how the quantized Hall conductance can be measured from
density profiles using the St{\v r}eda formula~\cite{Str1982}. In this paper 
we propose to measure the quantized Hall conductance directly as a transport property and suggest to use the response 
of the considered system to the time dependent perturbation. 

Before we turn to atomic systems, it is instructive to review briefly the phenomenology of 
electronic systems~\cite{yos,doucot}. Hall effect appears already in classical physics, 
where the transverse resistivity is proportional to the magnetic field $B$. 
The transverse conductivity may be expressed by the famous expression
\begin{equation}
\sigma_{yx}= \nu \frac{e^2}{h},
\label{formula}
\end{equation}
where $\nu=nh/eB$ is the filling factor, and $n$ is the electron  areal density. 
Amazingly the same formula holds in the quantum mechanical case, 
being the consequence of the Galilean invariance~\cite{doucot}.

The  explanation of plateaux corresponding to integer filling factors observed in experiments 
requires thus additional arguments. These arguments are based on the fact that 
in typical experimental situations Galilean invariance breaks down due to the presence of random impurities. 
Accordingly, the spectrum of (non-interacting) 2D electron gas in the magnetic field 
does not exhibit discrete Landau levels only. Close to the Landau levels in fact 
the spectrum consists of a band corresponding to extended (conducting) states. 
Far from the Landau level energy the spectrum corresponds to states localized due to the presence of 
impurities via the mechanism of  Anderson localization \cite{Anderson1958}.  Obviously, 
localized states do not contribute to the conductivity, and thus one can expect that when Fermi energy 
decreases between two subsequent Landau levels including less and less localized states in the Fermi sea, 
keeping the condition of fully occupied Landau levels, 
the conductivity does not change i.e. it exhibits a plateau. 
Why has the plateau the value exactly equal to $\sigma_{yx}= \nu \frac{e^2}{h}$ with $\nu$ integer, 
and why is this result so robust was a rather surprising fact in the beginning of the 1980s. 
It was first explained by the famous Laughlin argument~\cite{Laughlin1981}. He demonstrated the 
quantization of the Hall resistivity analyzing an imaginary experiment topologically equivalent to 
a "Corbino-type" sample of a disk shape with a central hole \cite{yos}.  Laughlin's arguments 
were then generalized by Halperin~\cite{Halperin1982} and B\"uttiker~\cite{Buttiker1988} to the strip geometry, 
employing the properties of the edge states and edge currents. Contemporary understanding of the robustness 
of the IQHE is based on the topological nature of integer transverse conductivity, 
first related to Chern numbers by Niu et al.~\cite{Niu1985} (see also~\cite{Wenbook} and references therein).  

The IQHE requires high, but not extensively high values of the magnetic field, 
in which several Landau levels are involved. The quantization of the transverse conductivity 
corresponding to integer values of the filling factor, is of course due to the quantization of the Landau levels.  
Still, the step-wise behavior of conductivity in the 2D electron gas originates from the 
influence of impurities~\cite{Laug1983}. 

In contrast, in the case of strong magnetic fields in the lowest Landau level regime, 
the non-interacting particle approach cannot be applied. One can think of the composite-fermion picture 
in which the fractional filling factor for electrons is transformed into integer filling factors for 
new quasiparticles: electrons dressed with magnetic quantum fluxes fill completely several 
Landau levels~\cite{Jain1989,Jainbook}. However, for this equivalent system of composite fermions in the IQH regime, 
the role of impurities is strongly combined with effects of interaction. One can think that 
the impurities play the role of a reservoir of particles trapping or releasing particles as the Fermi 
level moves across the localized states as the magnetic field changes. Notice that 
the Fermi energy decreases to lower values as the real magnetic field $B$ grows \cite{yos}. 
As a consequence, for some intervals of $B$ the density of the extended electrons, 
those that contribute to the current,  increases due to the transfer of electrons from impurities to the Landau levels. This effect compensates the increase of $B$, providing the appearance of a plateau in the resistivity/conductivity:

\begin{equation}
\label{eq:mainhyp}
\rho_{yx}\sim B/n_e,
\end{equation}
where $\,n_e\,$  is the density of the extended part of the system. It must 
be stressed that the presence of impurities plays the same role as that of the edge in finite systems. 
The main ingredient is the presence of a scalar potential locally linear in $\,x\,$ that traps the particles \cite{yos}.

The size of the plateaux depends thus on the number and the properties of the impurities. 
It must be realized that these plateaux appear on special values of $\rho_{yx}$ that localize states 
of significant interest, with fractional filling factors. Without impurities 
these values of the resistivity would not be visible. 

\medskip    

Turning back to the atomic gases, to simulate the similar phenomenology, we must somehow obtain 
localized and non-localized particles and look for regimes where transfer between them is possible. 
Since in ultracold atom setups one has the possibility to engineer controllable  impurities, such systems provide 
an ideal tool to understand the role of impurities in the formation of the plateaux and their interplay 
with interactions in the FQHE. In our numerical simulations, the possibility of distinguishing localized and 
non-localized particles was achieved in the following way: The diagonalization of the one-body density matrix 
provides us with the natural orbitals. 
Importantly, in all the analyzed cases, one of the orbitals is mostly concentrated around the impurity. 
In contrast, the other ones remain extended. 
Therefore we can distinguish between these two parts. Intervals of the artificial magnetic field $B^*$ 
where the occupations of the natural orbits have a significant variation with $B^*$ turned out to 
be crucial to identify the regions where transfer is possible and plateaux are expected.

Recently we have used state-of-art exact diagonalization to study properties of small clouds of atoms 
in a trap under influence of strong artificial gauge fields (see~\cite{Bar2006,Dag2009,Dag2007,JuliaDiaz2011,JuliaDiaz2012,Grass2012b,Grass2014a,Grass2014b}). 
In this paper, we expand the previous studies \cite{pin}. Our main goal is to learn 
about the relationship between the presence of impurities and the appearance of plateaux in 
the Hall resistivity as a function of the magnetic field in the fractional quantum Hall regime. 
This regime has not been achieved experimentally with atoms and our preliminary analysis is intended 
to predict possible future results. We use here the exact diagonalization method to 
calculate the ground state (GS) and its excitations in the absence/presence of an impurity. 
We analyze the time evolution (TE) of the system submitted to a periodic perturbation 
which represents an applied external electric field to simulate the transport equation 
\begin{equation}
\label{eq:transport}
j_y(t)=\sigma_{yx} E_x(t)	\,\,,
\end{equation}
where $j_y$ is the equivalent to the electronic current for atoms and $\sigma_{yx}$ is the transverse conductivity. 
In the Appendix we compare the results with those obtained using the linear response theory (LRT) approximation 
and conclude that, aside some limitations associated with resonances, the 
comparison is extremely good. 

Our main result is the appearance of a plateau close to the GS transition that takes place 
when the angular momentum changes from $L\,=\,N(N-2)\,$ to $L\,=\,N(N-1)$. The change of angular 
momentum modifies the resistivity and produces a bump partially overlapped with the plateau. 
An important outcome is that the presence of an impurity is a necessary condition to generate  plateaux. 
With no impurities the change of the occupations is abrupt and the transfer process is not  possible.  

The numerical complexity of the problem allows us to study only rather small systems up to $N=4$ atoms. In effect the
predicted plateau is small. 
We expect, however, that for large systems ($N=100$--$1000$) the natural increase of the number of impurities and thus the increase of the localized part, would guarantee the visibility of the plateau. On the other hand, the robustness comes from the topological nature of the conductivity, as previously mentioned.

This paper is organized as follows: 
In section II we present the model for the basic Hamiltonian $H_0$ 
to calculate the full spectrum of the unperturbed system. 
From its GS, we obtain the natural orbitals and their occupations. In addition, 
we show the explicit expression of the periodic perturbation and the expression for the equivalent periodic 
trapping potential in the full Hamiltonian. 
In section III, we analyze the time evolution of the expected value of the current operator $j_y$ and identify 
the conductivity. The main results are shown. In section IV we define a kind of restricted operators 
that emphasize the presence of the plateau and discuss about its meaning. 
Finally section V contains the summary and discussion. 
The appendix develops the LRT and shows some comparisons with the TE, 
as well as some limitations inherently related with the method.

\section{The model }

We consider a system of $N$ one-component bosonic atoms of mass $M$ confined on the $XY$-plane. The cloud
is trapped by a rotating parabolic potential of frequency $\omega_{\perp}$ and
rotation $\Omega$ along the $Z$-axis, rotation which in an effective way generates an artificial magnetic field denoted by $B^*$. In the rotating reference frame 
the basic Hamiltonian  (not including the perturbation) reads
\begin{equation}
\hat{H_0}\,\,\,=\,\,\, \hat{H}_{\rm sp}+\hat{H}_{\rm{int}},
\end{equation}
where the single particle part is given by
\begin{equation}
\hat{H}_{\rm{sp}}= \sum_{i=1}^N\bigg[\frac{\hat{\bf p}^2}{2M}+\frac{M}{2}\omega_{\perp}^2\hat{\bf r}^2 - \Omega {\bf z}\cdot \hat{\bf L}\bigg]_i + \hat{W}
\label{Ham_SP}
\end{equation}  
which can be rewritten in an equivalent way as
\begin{eqnarray}
\label{eq:Hamnopert}
\hat{H}_{\rm{sp}} & = & \sum_{i=1}^N\bigg[\frac{1}{2M}(\hat{\bf p}+\hat{\bf A} )^2 + \frac{1}{2}M\left( \omega_{\perp}^2-\frac{(B^*)^2}{4M^2}\right)\hat{\bf r}^2\bigg]_i+ \hat{W}\,,
\end{eqnarray}
with
\begin{equation}
\hat{A}_x=\frac{B^*}{2}\hat{y}\,\,\,,\,\,\,\hat{A}_y=-\frac{B^*}{2}\hat{x}\,.
\end{equation}
The particular selection of the symmetric gauge has
been done in the definition of $\hat{\bf A}$, with 
\begin{equation}
\label{eq:Brotation}
B^*=2M\Omega 
\end{equation}
being the modulus of a
constant artificial magnetic field directed downward along the
$Z$-direction and $\bf r \equiv (x,y)$. 

The potential $\hat{W}=\sum_{j=1}^K \hat{W}_j$ is due to the presence of  $K$ impurities, which are modeled by Dirac delta functions, 
\begin{equation}
\hat{W}_j=-\gamma_j \frac{\hbar^2}{M} \sum_{i=1}^N \delta^{(2)}(\hat{\bf r}_i- \bf a_j)
\label{defect}
\end{equation}
The dimensionless parameters $\,\gamma_j\,$ measure the strength of the impurities and $\bf a_j$ localize them on the $XY$ plane. The term $\hat{W}$ breaks the circular symmetry except for the case of a single impurity localized exactly at the center.

We model the atomic interaction by a 2D contact potential characterized by
\begin{equation}
\hat{H}_{\rm{int}}=\frac{\hbar^2 g}{M}\sum_{i<j} \delta^{(2)}(\hat{\bf r}_i-\hat{\bf r}_j)\,,
\end{equation}
where $g=\sqrt{8\pi}a/\lambda_z$ is the dimensionless coupling, $a$ is the 3D scattering length and 
$\lambda_z=\sqrt{\hbar/M\omega_z}$. We assume $\omega_z$, the trap frequency in the $Z$-direction, 
much larger than any of the energy scales involved, in such a way that only the lowest 
level is occupied. Therefore, the dynamic is frozen in the $Z$-axis and the system can be considered two-dimensional.

Without impurities, the solutions of the single particle part produce the Landau level structure \cite{Jainbook}. 
The energy levels are separated by $\hbar (\omega_{\perp}+\Omega)$. We assume that $B^*$ is large enough 
to consider just the lowest Landau level (LLL) regime where the appearance of energy gaps has a 
completely different origin as those in the IQH, where several Landau levels are implied. 
Within this regime, the kinetic part of the Hamiltonian reads
\begin{equation}
\hat{H}_{\rm{kin}}=\hbar(\omega_{\perp}-\Omega)\,\hat{L}\,+ \hat{N}\hbar\omega_{\perp}\,.
\end{equation}
The single particle solutions with well defined angular momentum $m$ are the Fock-Darwin (FD) functions, given by 
$\,\phi_m(\theta,r) \,=\,\frac{e^{im\theta}}{\sqrt{\pi m!}} \,e^{-r^2/2}\,
r^m\,$.

\begin{figure}
 \begin{tabular}{ccccc}
 \hspace{-0.5cm}
\includegraphics[width=2.9cm]{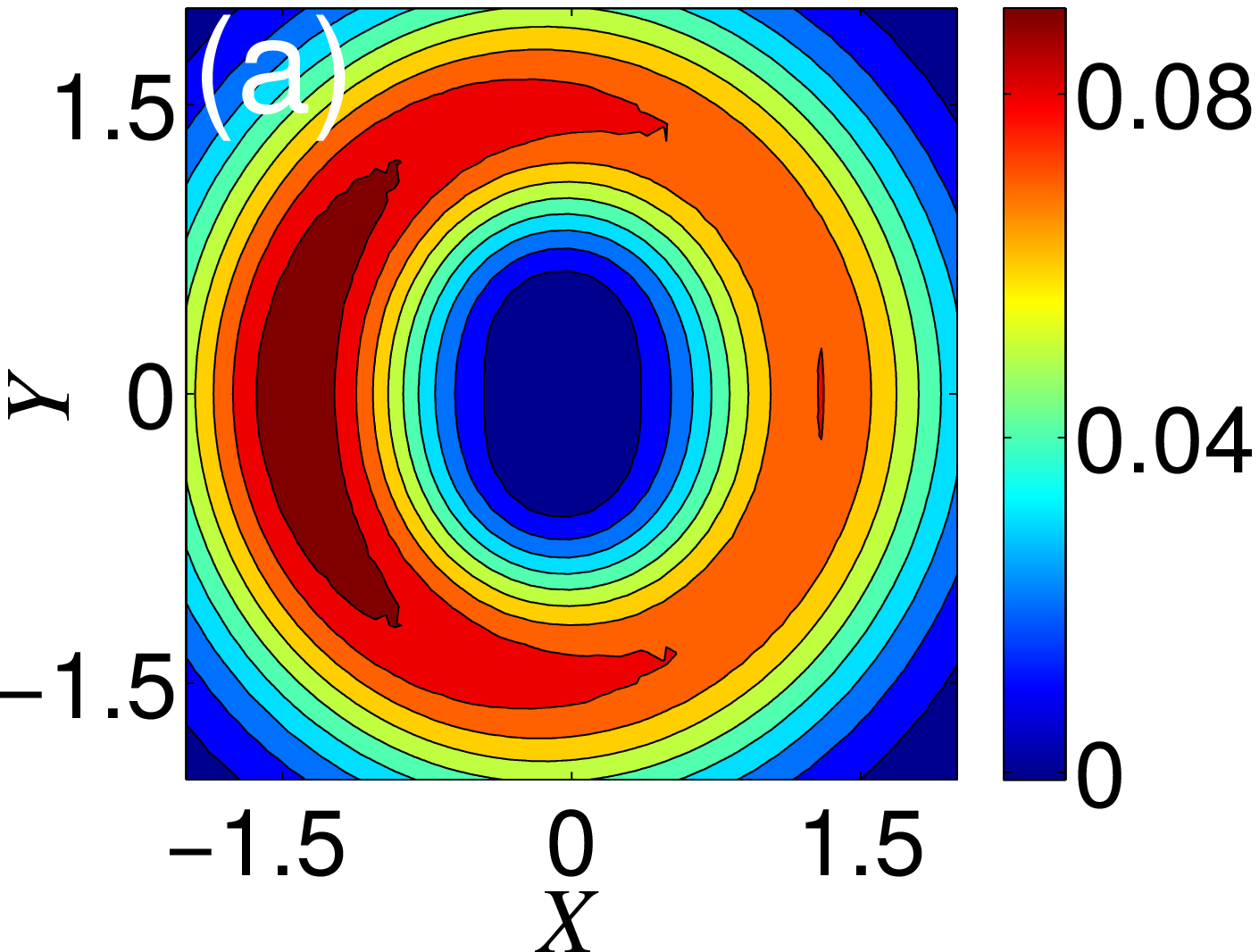}&\hspace{-0.5cm}
\includegraphics[width=2.9cm]{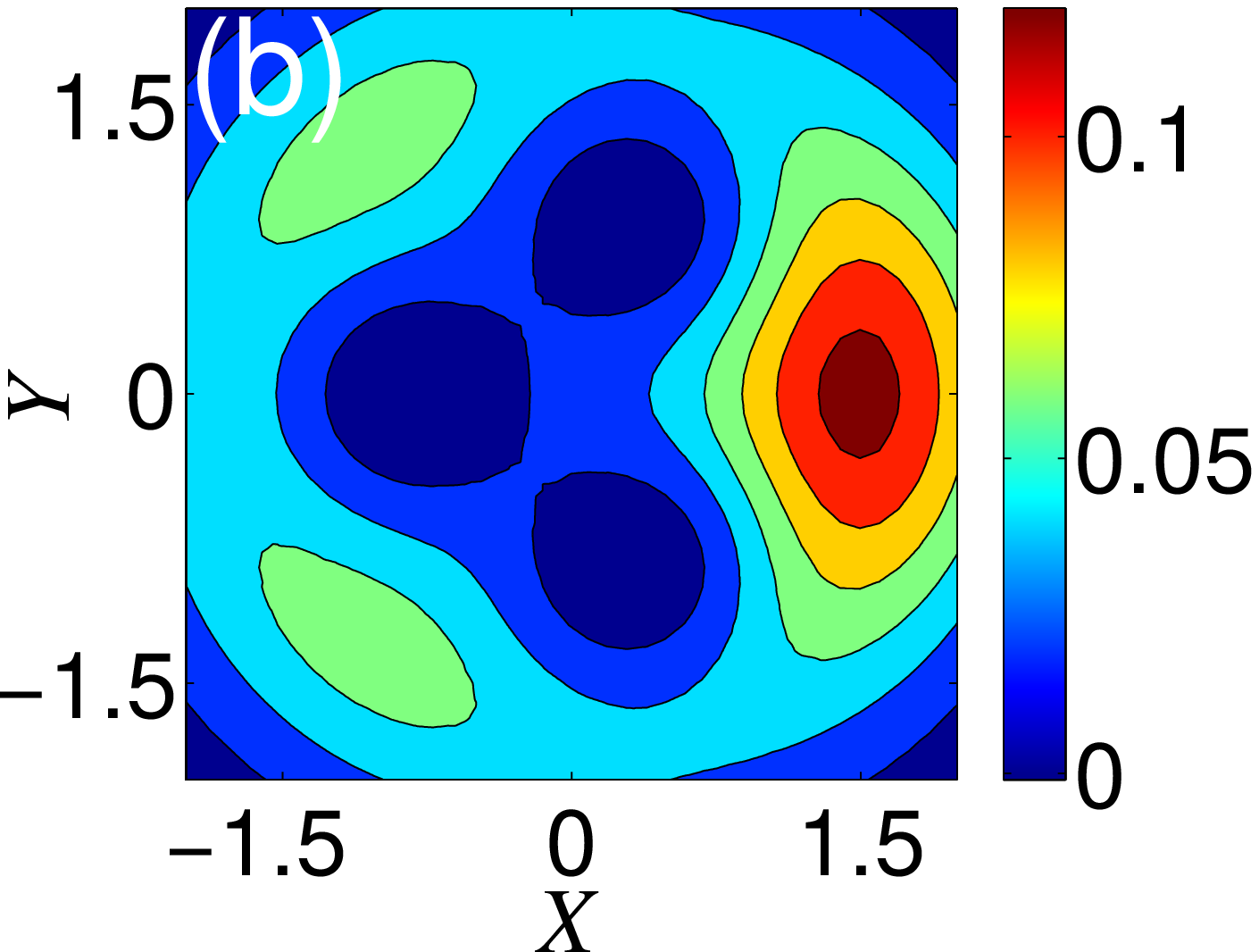}&\hspace{-0.5cm}
\includegraphics[width=2.9cm]{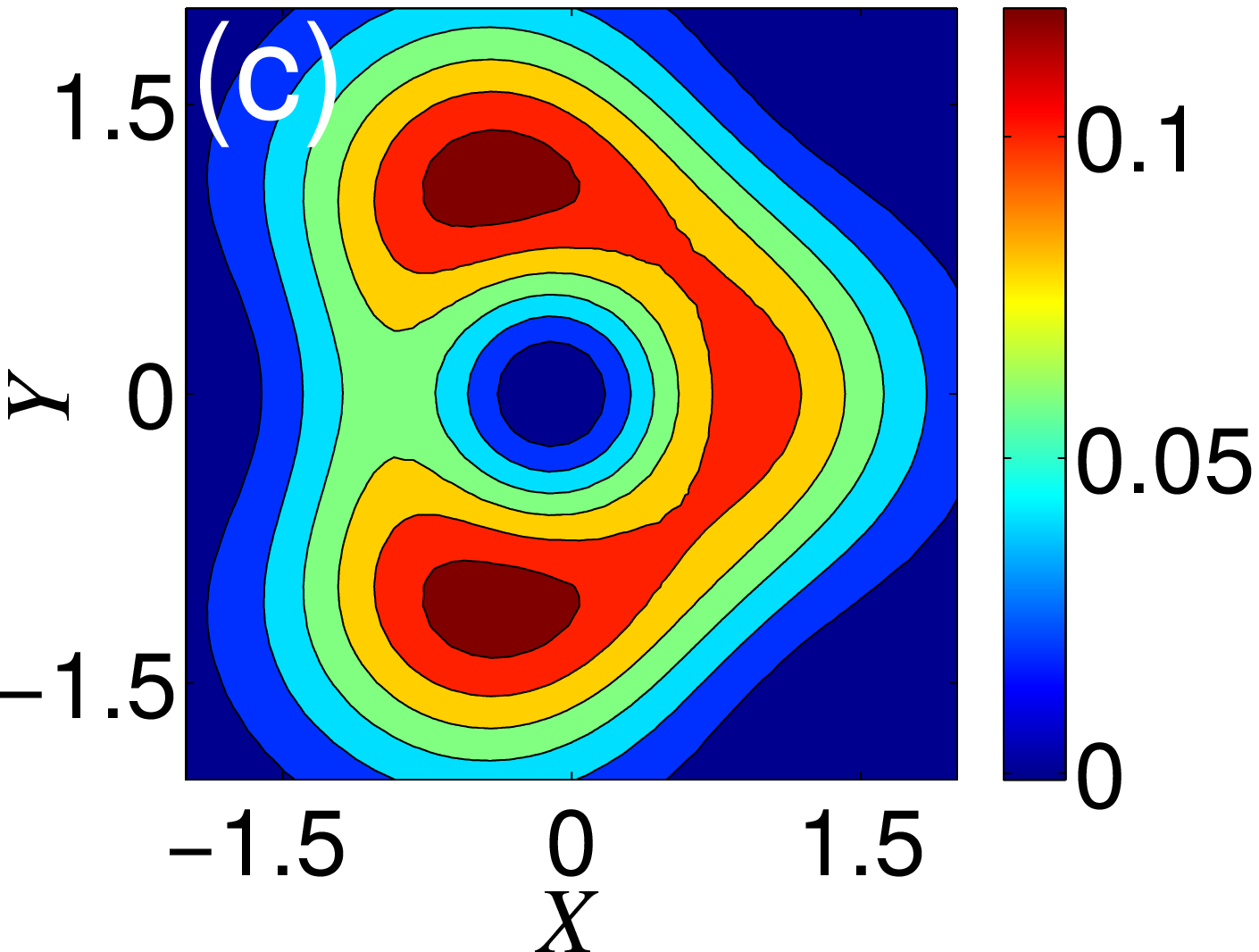}&\hspace{-0.5cm}
\includegraphics[width=2.85cm]{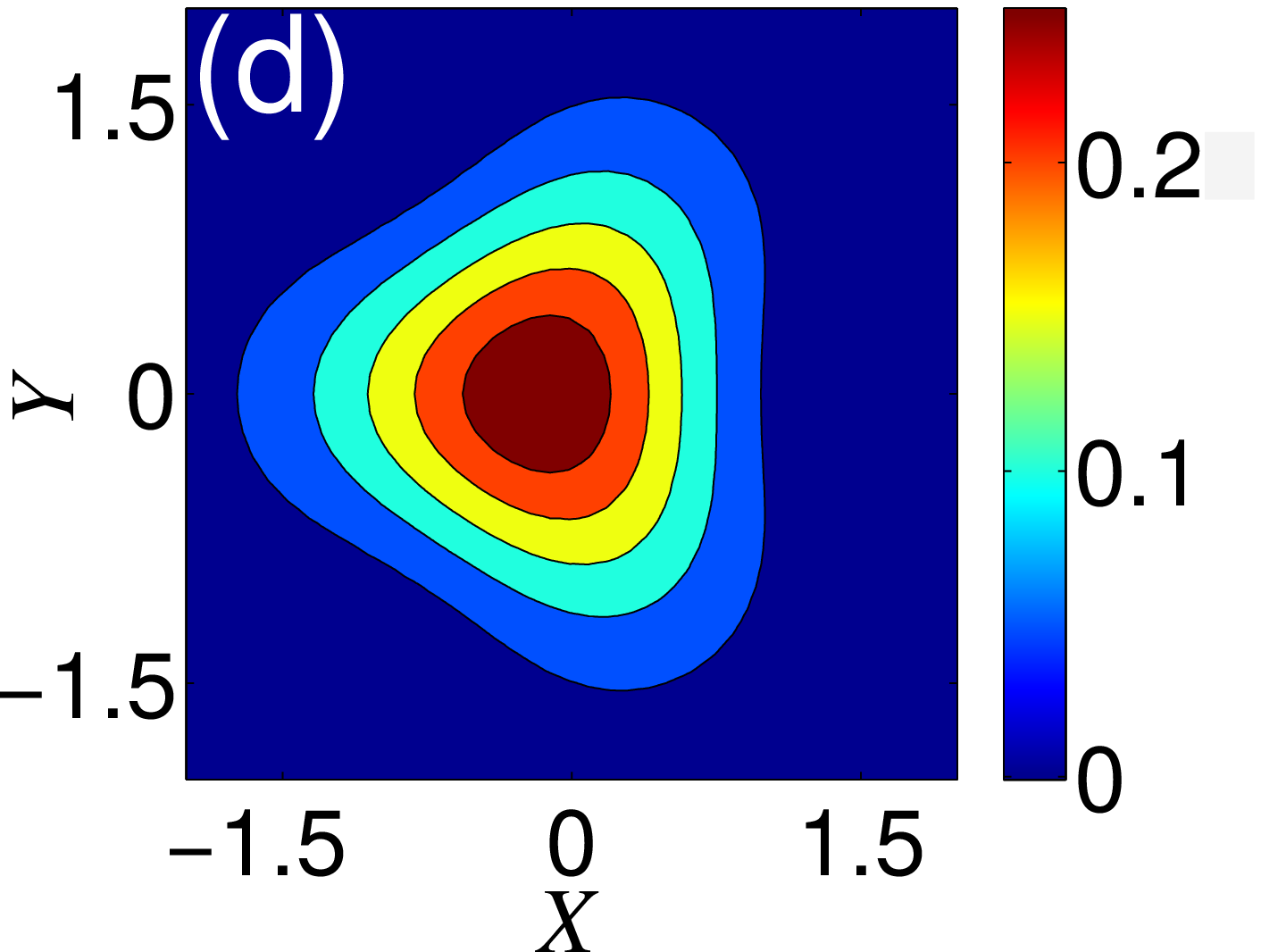}&\hspace{-0.5cm}
\includegraphics[width=2.95cm]{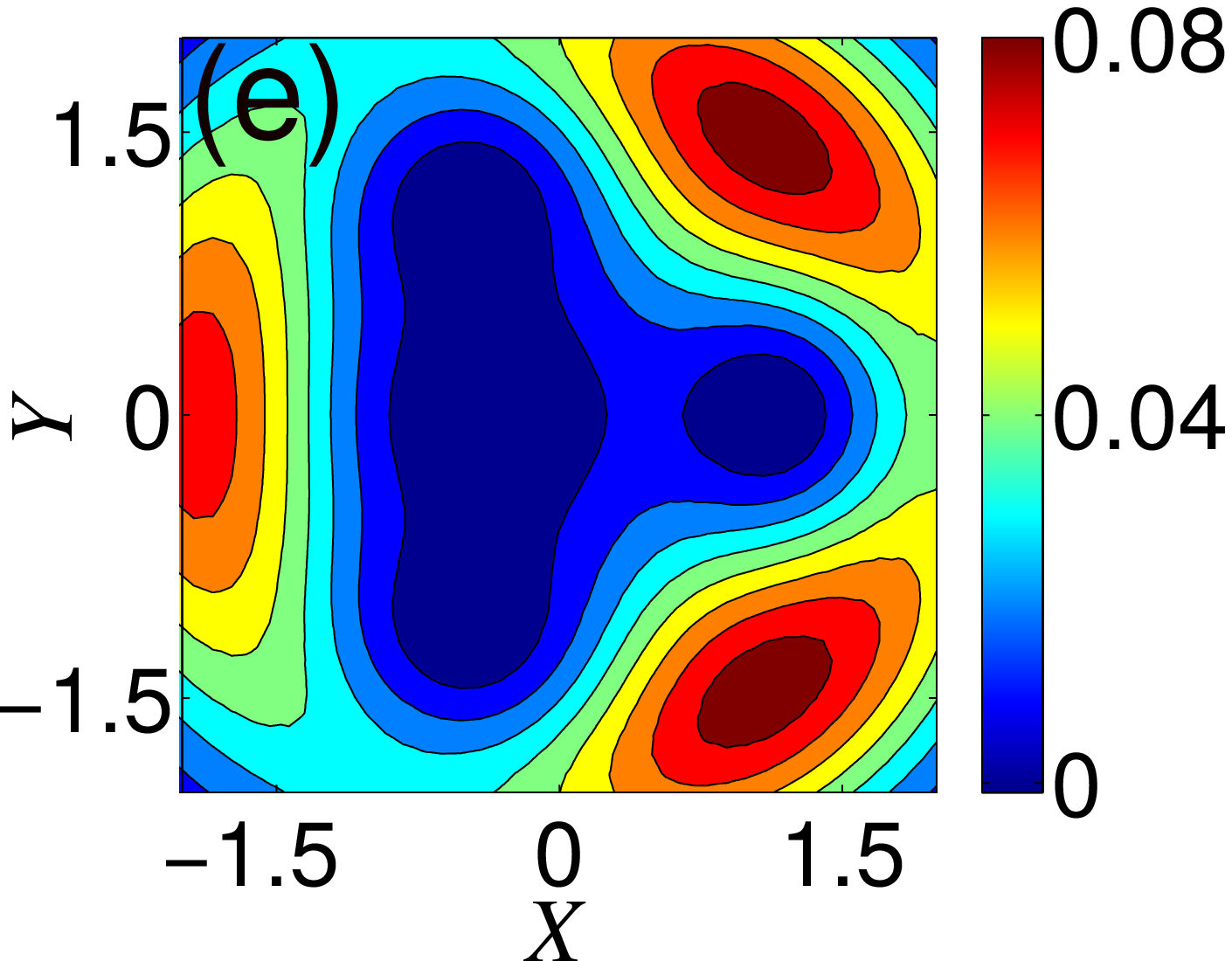}\\
 \hspace{-0.5cm}
\includegraphics[width=2.9cm]{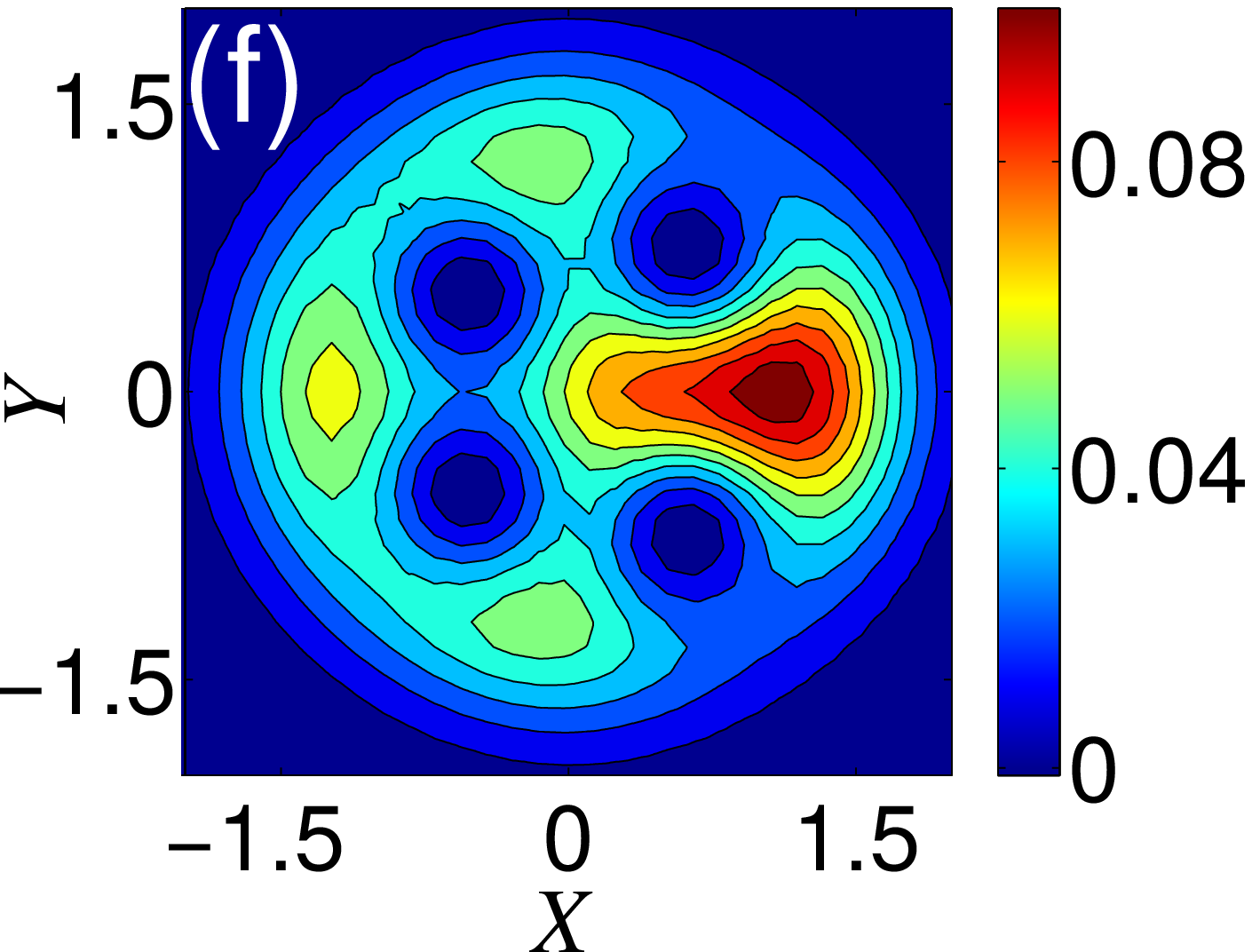}&\hspace{-0.5cm}
\includegraphics[width=2.9cm]{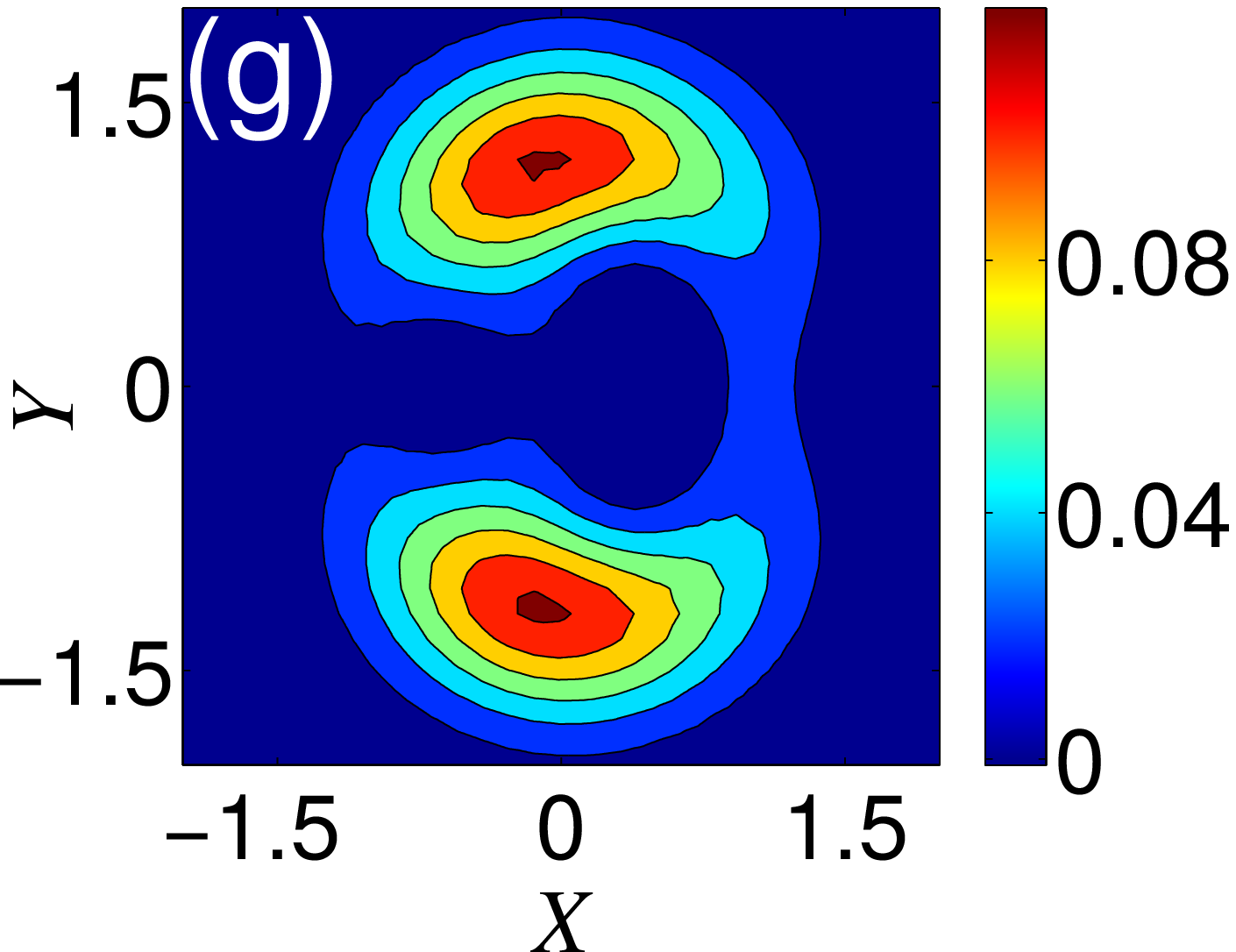}&\hspace{-0.5cm}
\includegraphics[width=2.85cm]{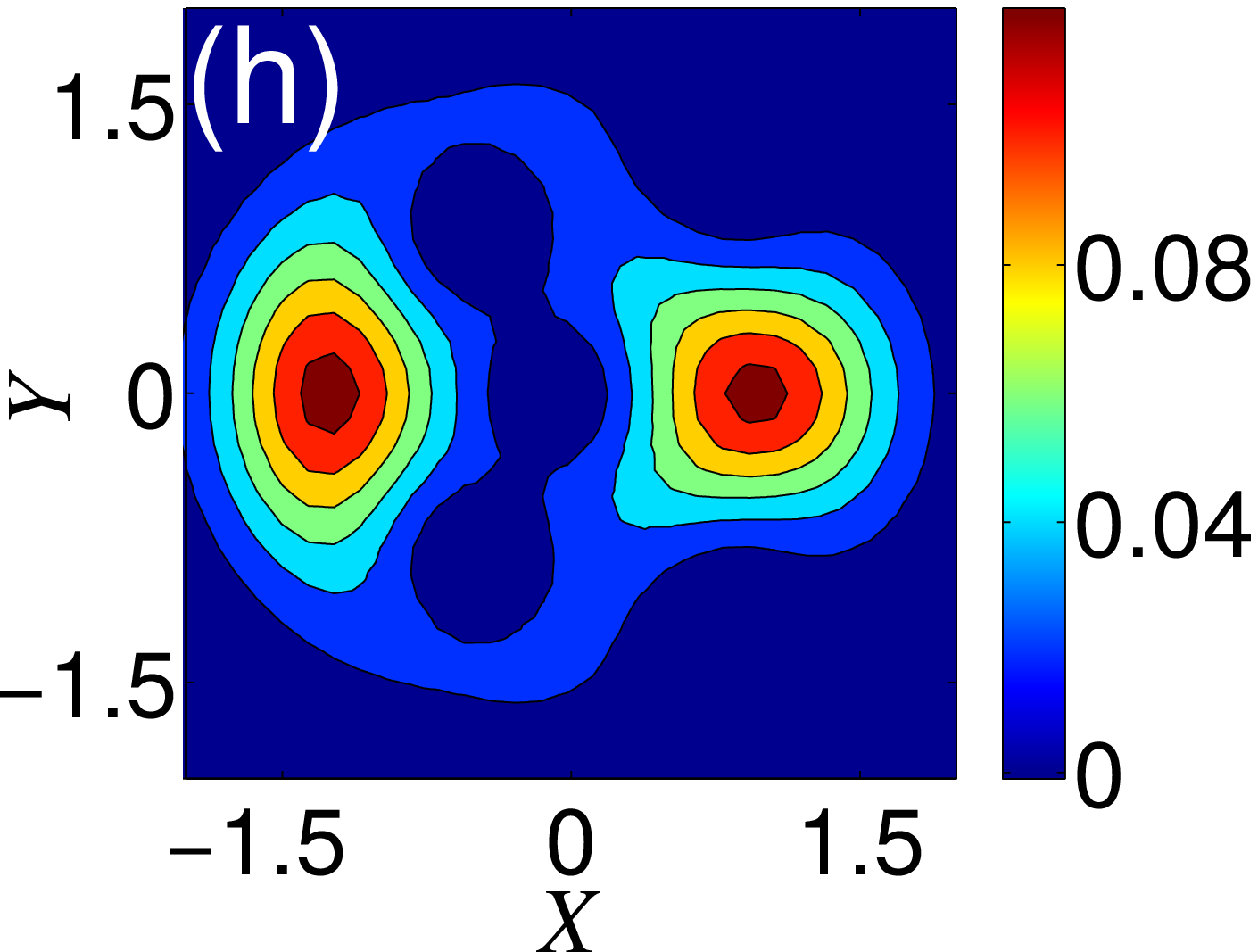}&\hspace{-0.5cm}
\includegraphics[width=2.9cm]{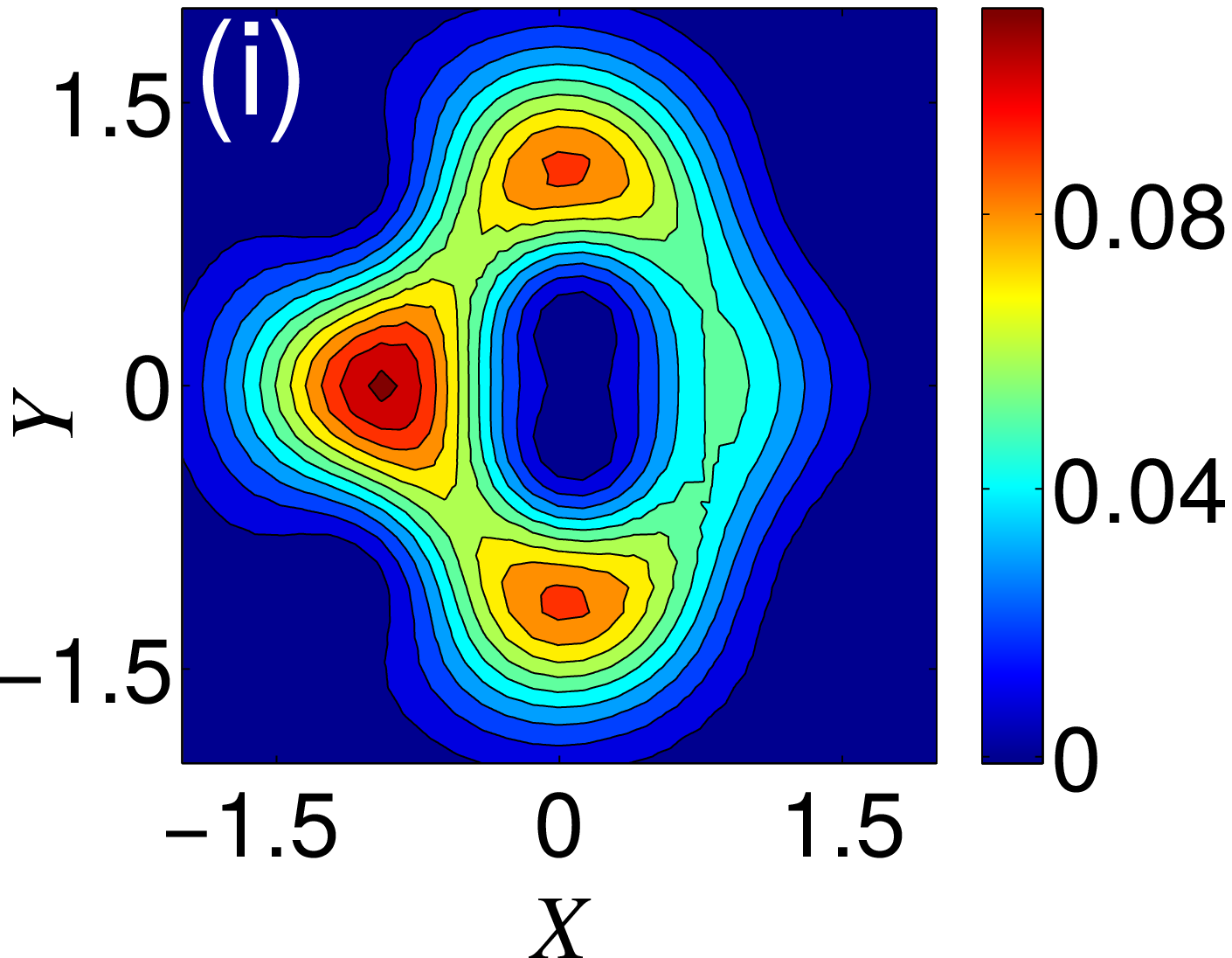}&\hspace{-0.5cm}
\includegraphics[width=2.95cm]{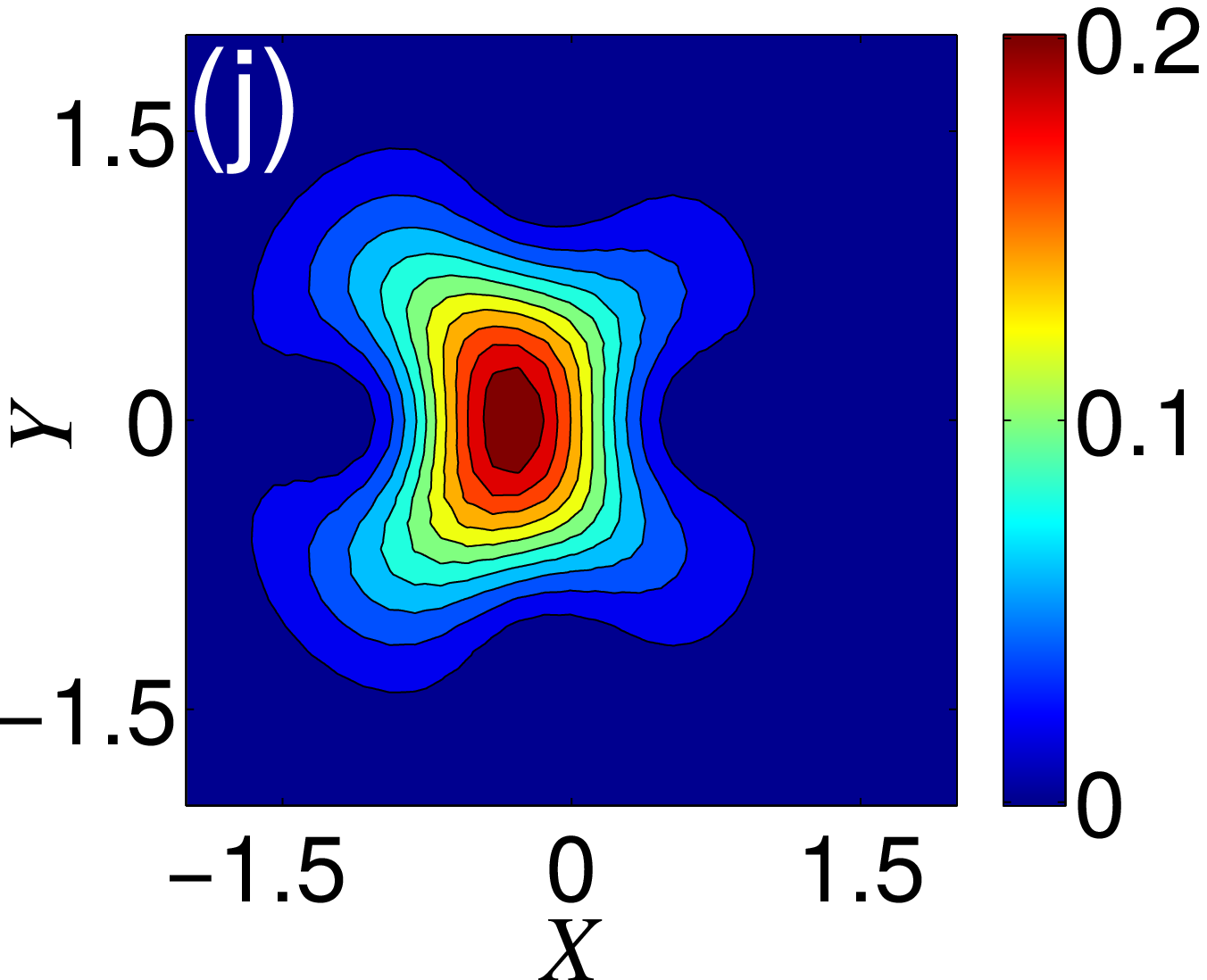}
 \end{tabular}
\caption{(Color online) Density of the first five natural orbitals, the only ones with non-negligible occupations for $N=3$ (upper row) and for $N=4$ (lower row). For $N=3$, the second one, panel (b), is localized close to the impurity siting at $\bf a_1=(1,0)$.  In contrast, for  $N=4$, the natural orbital which is localized around the impurity is the first one [panel (f)]. The occupations are:  $\,0.92\,,0.74\,,0.57\,,0.42$ and $0.34$
for $N=3\,$ and $1.30\,, 0.83\,, 0.68\,, 0.52\,, 0.41$ and $0.26$ for $N=4$, respectively. In both cases we take  $\,Ng=6\,$, $\,\gamma_1=0.1\,$, $\,\Omega=1.95\,$. \label{fig1}}
\end{figure}
\medskip

Once the spectrum of the whole system is obtained for a given number of particles $N$ 
and for fixed values of $\Omega,\,\, \gamma_j, \,\,\bf a_j\,\,$ and $\,\,g$, 
we proceed to distinguish 
the eigenfunctions of the one-body density matrix 
according to their localization. To this end, we diagonalize the one-body density matrix given by
\begin{equation}
\label{eq:densM}
\hat{\rho}^{(1)} (\bf r,\bf r')\,=\,\langle \hat{\Psi}^{\dag}(\bf r) \hat{\Psi}(\bf r')\rangle,
\end{equation}
where the expected value is calculated at the GS and $\hat{\Psi}(\bf r)$ is the field operator.  
Its eigenfunctions are the natural orbitals $\,{\psi_i}\,$, linear combinations of the FD functions, 
and the eigenvalues are their occupations $\,{n_i}\,$, $i=1,..,l_m+1$ ($l_m$ varied until convergence). 

From now on, we use the complete set of natural orbitals as a base to represent functions and 
operators in the second quantized formalism. Next we analyze the density distribution of each orbital 
and look for their localization around the impurities. For $\,N=3\,$ and one impurity at 
$\bf a_1=(1,0)$ [see Eq.~(\ref{defect}) - lengths are in units of the $XY$ harmonic oscillator length],  
the result is that the orbital $\psi_2$ 
out of five orbitals with non-negligible occupation (the orbitals are ordered by decreasing occupations), 
presents a density distribution mainly localized at the impurity, as is shown in Fig.~\ref{fig1} (upper row). 
Similar results are obtained for $N=4$ where the most localized orbital is $\psi_1$ (lower row). 
This localization of some orbitals allows us to distinguish between localized and extended states.

Finally, we consider that the cloud of atoms is dynamically forced by an oscillating term, 
while the impurity remains attached to a fixed position. To be explicit, the full Hamiltonian of the system is 
\begin{equation}
\label{eq:totHam}
\hat{H}(t)=\hat{H}_0 + \hat{H}_{\rm pert}(t)
\end{equation}
with
\begin{equation}
\label{eq:pertH}
\hat{H}_{\rm pert}(t)=-\lambda\left(\sum_{i=1}^N \hat{x}_i\right) \xi(t) \sin(\omega t)\,\,\equiv \,\,
\sum_{i=1}^N f(t) \, \hat{x}_i,
\end{equation}
where $\lambda$ gives the intensity of the perturbation, which we assume small. The explicit form of $\xi(t)$ is 
\begin{equation}
\label{eq:xi}
\xi(t)=1-\exp[-(t/\sigma)^2]
\end{equation}
where $\sigma$ determines the velocity of the evolution. The perturbation is switched on at $t=0$ and, 
as $t$ increases, the stationary regime is achieved when the amplitude of the oscillations can be considered  constant. 
From now on we consider $M=1/2$
and $\hbar=1$ and choose 
$\,\lambda_{\perp}=\sqrt{\frac{\hbar}{M\omega_{\perp}}}=\sqrt{2/\omega_{\perp}}\,$, $\,\hbar\omega_{\perp}/2\,$ 
and $\,\omega_{\perp}/2\,$ as units of length, energy, and frequency, respectively. With our unit of length, 
$\,\omega_{\perp}=2$. In the simulation, the sequence 
$-\lambda \xi(t) \sin(\omega t)$ is identified with the electric field $E_x(t)$ in the transport equation. Namely, 
for a single particle and a single impurity [see Eq.~(\ref{Ham_SP})], including the perturbation we have
\begin{equation}
\hat{H}_i= (\hat{\bf p}+\hat{\bf A})^2_i + (\omega_{\perp}^2-\Omega^2)(\hat{\bf x}^2+\hat{\bf y}^2)_i + f(t)\hat{\bf x}_i -\gamma_1 \delta^{(2)}( \hat{\bf r}_i-\bf a_1).
\end{equation}
The effective trapping potential can be re-written as an oscillating trap
\begin{equation}
(\omega_{\perp}^2-\Omega^2)\bigg[(\hat{\bf x}+\frac{f(t)}{2(\omega_{\perp}^2-\Omega^2)})^2+\hat{\bf y}^2\bigg]_i.
\end{equation}

\section{Time evolution}

\begin{figure}[tbp]
 \begin{tabular}{cc}
 \includegraphics*[width=0.47\columnwidth]{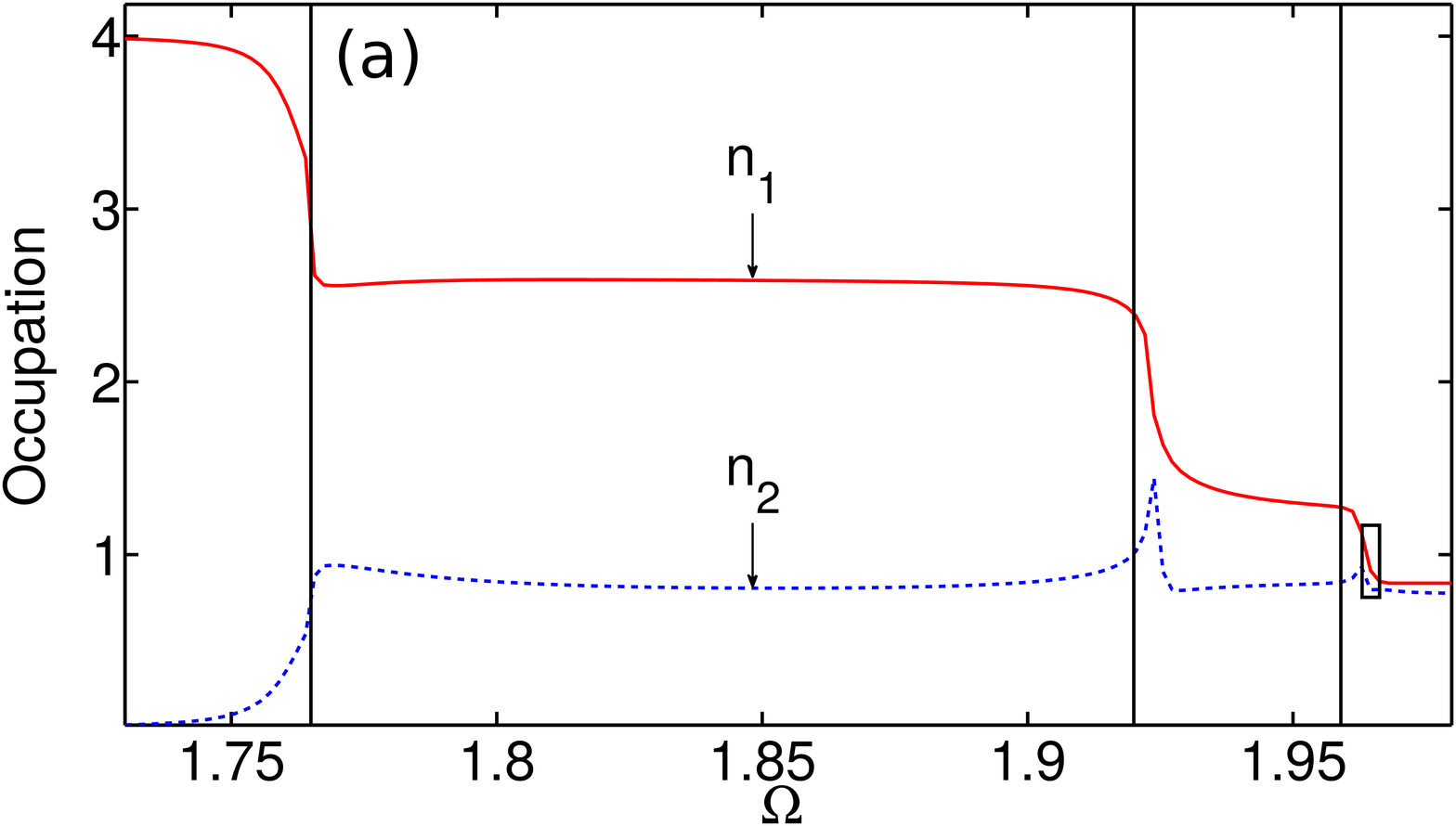}&\includegraphics*[width=0.47\columnwidth]{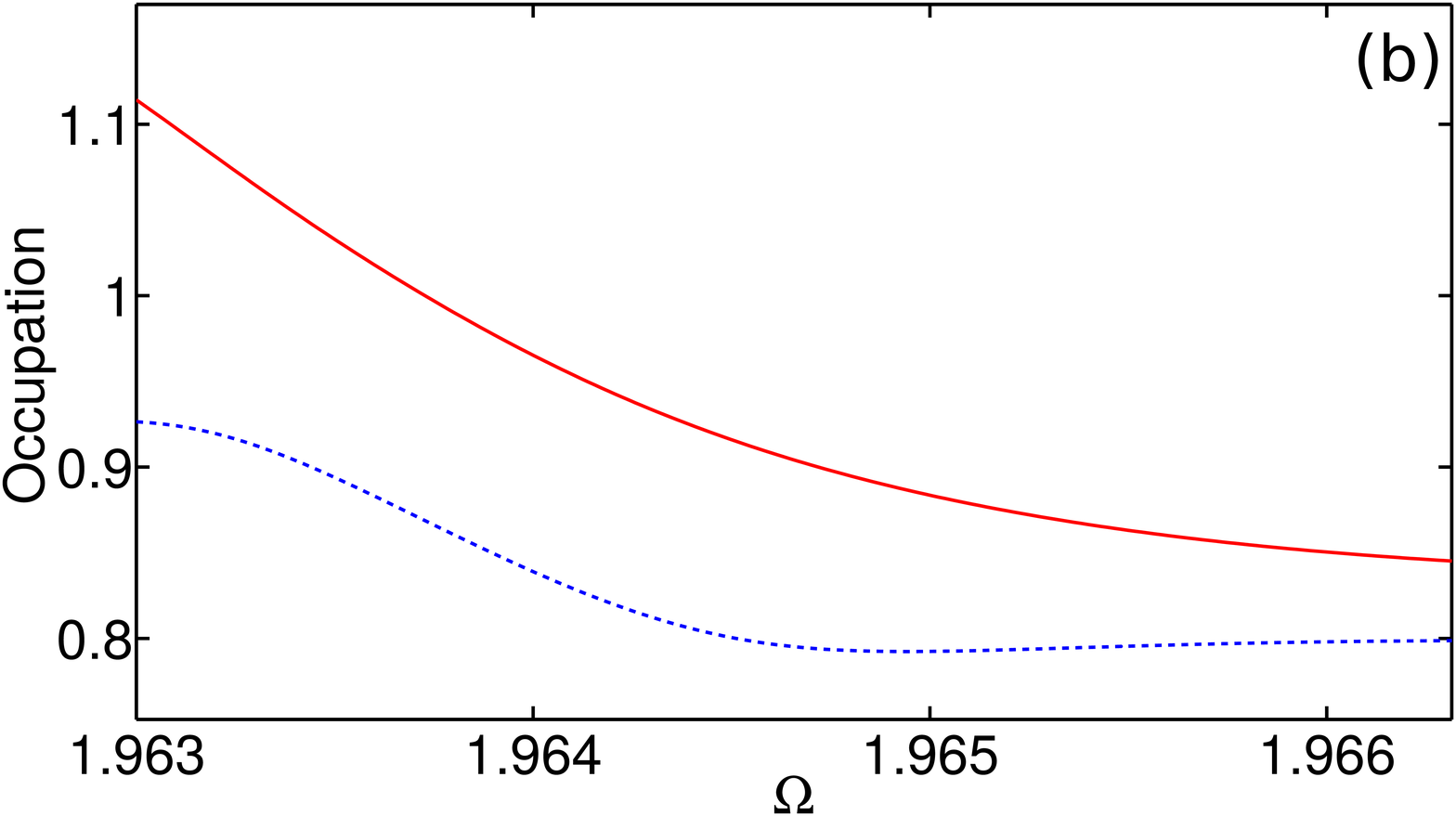}
 \end{tabular}
\caption{(a) Occupations of the first two orbitals as a function of $\Omega$. There is a correlation between the steps of the occupations and the jumps of $\langle L \rangle$ in a nearly symmetric system. For $N=4$ the magic numbers of $L$ in the symmetric case are $0-4-8-12$. The vertical lines at $\,1.765\,\,,1.920\,\,$ and $\,1.959\,$ mark the position of the jumps of $L$ for the symmetric case. We considered  $N=4\,$, $\,Ng=6\,$, $\,\omega=1.1\,$ $ \,\gamma_1=0.1\,$ and $\,\bf a_1=(1,0)$. (b) A zoom in the interesting region [which is highlighted with a small rectangle in (a)]. 
The localized orbital for $N=4$ is $\psi_1$.} \label{fig2}
\end{figure}

Let us show that this model allows us
to identify the transverse conductivity from the transport equation~(\ref{eq:transport}). For that, we need to analyze the time evolution of the expected value of the current operator $\hat{j}_y$, which
is given by~\cite{edm}
\begin{eqnarray}
\hat{j}_y(\bf r) & = & \frac{1}{2M}\{\hat{\Psi}^{\dag}(\bf r)\left[\hat{p}_y+\hat{A}_y(\bf r)\right]\hat{\Psi}(\bf r)
\nonumber
\\
& - & \left[\left[\hat{p}_y-\hat{A}_y(\bf r)\right]\hat{\Psi}^{\dag}(\bf r)\right]\hat{\Psi}(\bf r)\}.
\label{eq:jy}
\end{eqnarray}
By calculating $\langle \hat{j}_y \rangle_t$ once the stationary regime is reached and {\it in the case } 
that we obtain a linear behavior in $\lambda$ [see Eq.~(\ref{eq:pertH})], 
we 
are able to obtain the transverse conductivity from the transport equation~(\ref{eq:transport}) 
due to the identification we have done between the perturbation and the electric potential 
associated with a constant electric field.  

To obtain $\langle \Psi(t)|\hat{j}_y |\Psi(t)\rangle$ we solve the Schr\"{o}dinger equation 
$\,\, i\partial_t \Psi(t)=\hat{H}(t) \Psi(t)\,\,$ with the time-dependent Hamiltonian given by Eq.~(\ref{eq:totHam}). 
We consider the wave function as $\Psi(t)=\sum_{n=1}^{n_d}\,c_n(t)\, \Phi_n$ 
where the set $\{\Phi_n\}$, $n=1,...,n_d$ is a basis of the Hilbert space of dimension $n_d$, 
given by the many-body wave functions which solve Hamiltonian $H_0$ with eigenvalues $\{E_n\}$. 
Then, we obtain the system of equations:
\medskip
\begin{eqnarray}
i \partial_t\,{c}_n(t) &=&  c_n(t) E_n 
\nonumber
\\
& - & \,\lambda \,\xi(t) sin(\omega t) \,\,\sum_{m=1}^{n_d} c_m(t) <n|\hat{x}|m>		
\end{eqnarray}
with the GS$(\Phi_1)$ as the initial condition, i.e., $\Psi(t=0)\,=\,\Phi_1$. 
We solve these equations using the Runge-Kutta fourth-order algorithm. 
Once we obtain the transverse conductivity $\sigma_{yx}$, we can obtain the resistivity from 
\begin{equation}
\label{eq:resistivity}
\rho_{yx}=-\frac{\sigma_{yx}}{|\sigma_{yx}|^2+|\sigma_{xx}|^2},
\end{equation}
where $\sigma_{xx}$ is obtained from $\,j_x=\sigma_{xx} E_x\,$.

Let us note here that our evolution is rather adiabatic. Indeed, 
we have checked that the overlap $|\langle \Psi_{GS}|\Psi(t)\rangle|$ remains nearly one at all times. 
E.g., we numerically obtain that, for $\sigma=10$ (see Eq.~(\ref{eq:xi})), 
\begin{equation}
|\langle \Psi_{GS}|\Psi(t_c)\rangle|^2=0.97
\end{equation}
where $\Psi_{GS}$ is the ground state of $\,\,\hat{H}(t_c)=\hat{H_0}+\hat{H}_{pert}(t_c)\,\,$ and $\,\Psi(t_c)\,$ is the solution of the Schr\"{o}dinger equation at $t_c$ (chosen such that $sin(\omega t_c)=1$).  

In agreement with our previous discussion, to generate a plateau we must look for an interval of $\Omega$ 
where the occupation of the localized orbital changes. To this end, we analyze the orbital occupations 
as a function of $\Omega$  [see Fig.~\ref{fig2} (a)] and focus on the region fulfilling two requirements: 
On one hand, the occupation of the localized orbital decreases as $B^*$ increases producing an increase 
of the extended part and, on the other hand, this region lies within the largest possible value of 
$\langle L \rangle$ where plateaux are expected. In other words, there are two kind of intervals: 
i) intervals where the localized density is flat giving linear dependence of $\rho_{yx}$ with $B^*$ 
where quantum and classical behavior coincide \cite{yos}; and ii) intervals where a plateau occurs and then, 
the change of $B^*$ drags $\rho_d$ (the extended density), and $\rho_{yx}$ remains constant. 
Notice that this means that for our analyzed small samples we expect only one plateau along the whole interval 
of the largest value of $\langle L \rangle$. Fig.~\ref{fig2} (b) is a zoom of the interesting region.  



\begin{figure}[tbp]
\begin{tabular}{cc}
\includegraphics[width=0.47\columnwidth]{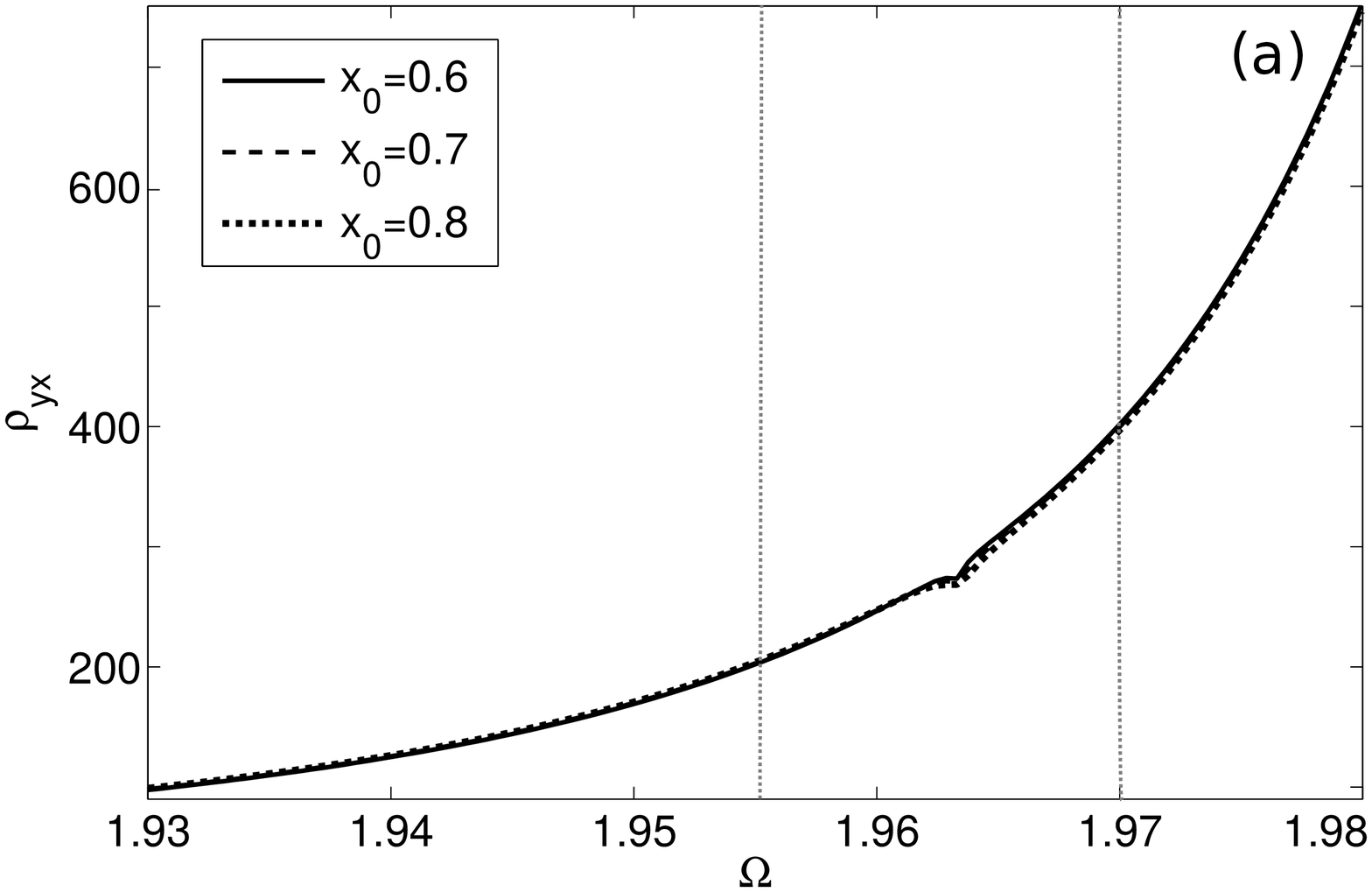} &
\includegraphics[width=0.47\columnwidth]{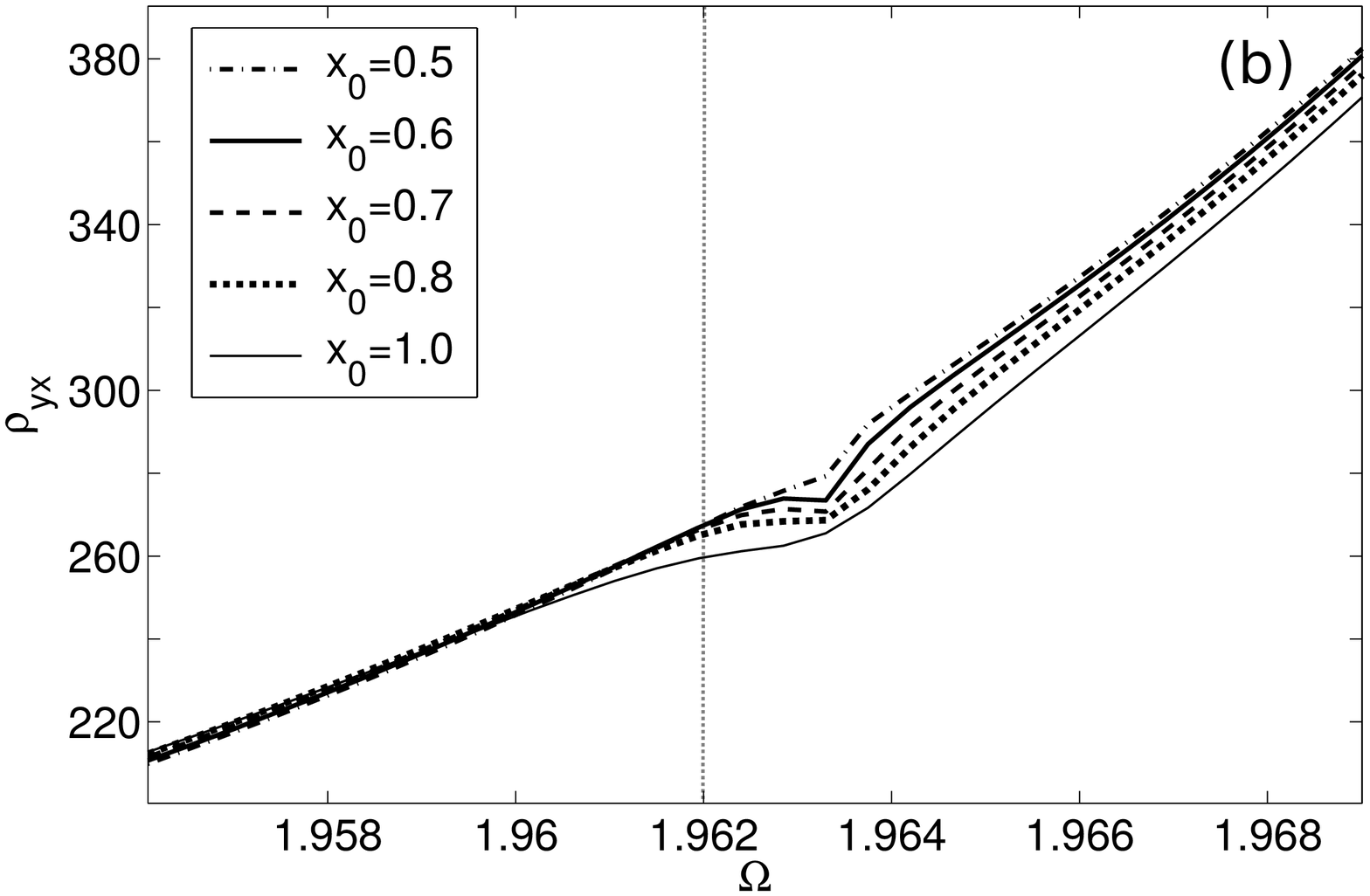}
\end{tabular}
\caption{(a) Hall resistivity $\rho_{yx}$ as a function of $\Omega$.  We used  $N=4$, $Ng=6$, $\omega=1.1$, $\gamma_1=0.1$, $\lambda=0.001$. We considered an impurity at the $X$-axis at different positions, that is, $\bf a_1=(x_0,0)$.  The vertical  dashed lines  mark the interval considered in panel (b), where we show a zoom of the area of interest. The vertical  dotted line at $1.962$ in panel (b) marks the approximate place of the jump in the angular momentum from 8 to 12. We also include two additional positions of the impurity to show that the effect depends on the position of the impurity. } \label{fig3}
\end{figure}

Our main result is represented in Fig.~\ref{fig3}, where we show the variation of the Hall resistivity as a function of $\Omega$ for some values of the position of the impurity, which we always assume to be in the $X$-axis, that is $\bf a_1=(x_0,0)$.  In Fig.~\ref{fig3} (a) we show the Hall resistivity for three exemplary values of the position of the impurity for the whole range of $\Omega$ investigated. We observe that  some structure, different from the linear behavior,  occurs just in the region shown in Fig.~\ref{fig2} (b).  This structure begins at about $\Omega=1.963$, that is just after $\Omega_c=1.9629$, where the jump of the angular momentum from $L=8$ to $L=12$ takes place. In Fig.~\ref{fig3} (b) we show a zoom of the relevant part of the figure, where we confirm that this structure corresponds to the expected plateau.  The total jump in the resistivity is for $x_0=0.7$ approximately from  $\rho_{yx}=270$ to 291, which is about a 7$\%$ of the total value of the resistivity. We also see that this plateau is soften when the position of the impurity is changed.

\begin{figure}[tbp]
\centering
\includegraphics[width=0.5\columnwidth]{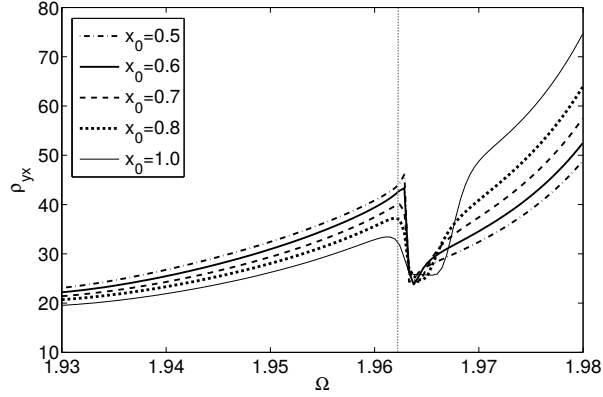}
\caption{Hall resistivity $\rho_{yx}$ as a function of $\Omega$ when restricting both the current and the position operators (see text). Same parameters as in Fig.~\ref{fig3}. The presence of a plateau is very evident when using the restricted operators. The plateau begins at about $\Omega=1.963$  just after $\Omega_c=1.962$ where the jump of the angular momentum from $L=8$ to $L=12$ approximately takes place (marked with a vertical gray dotted line).} \label{fig4}
\end{figure}

\section{Emphasizing the plateau structure} 

Up to this point, it has been proved that the most successful way to generate 
a plateau is by the localization of a very special interval of $\Omega$ where the localized density 
decreases (see Fig.~\ref{fig2}). Within this interval, there is a simultaneous change of 
$\,B^*\,$ and $\,\rho_d\,$. The necessary condition to produce this effect is 
under the requirement that for each $\,B^*\,$, its "dancing pair", $\,\rho_d\,$, is well defined. 
To be more precise: For each $B^*$, the deduction of $\rho_{yx}$ requires several steps, 
some of them are operations like $\sum_{qp} <q|\hat{j}_y|p>$ or $\sum_{qp}<q|\hat{x}|p>$ where $|p>$ and $|q>$ 
are Fock states: $|n_1,n_2,...n_{l_m}\rangle$, $n_i$ being the orbital occupations. 
In order to have a fixed value of $\rho_d$ the operators $\hat{j}_y$ and $\hat{x}$ 
must not do significant changes on the localized part.

To be sure that this phenomenology is well captured by our model, we forced the mechanism in 
an alternative calculation defining a kind of restricted operators $\tilde{j}_y$ and $\tilde{x}$ 
that fulfill the previous requirement: They do not change the localized part of the density. 
We consider in the summations only those elements that couple the vectors with the same occupation 
of the localized orbital (which is $n_2$ for $N=3$ or $n_1$ for $N=4$). 
The result shown in Fig.~\ref{fig4} confirms our intuition. Fig.~\ref{fig4} must be compared with Fig.~\ref{fig3}(a). 
The improvement of the visualization of the plateau for some values of $x_0$ is evident.  

Within the intervals of linear dependence of $\rho_{yx}$ with $B^*$ where quantum and classical behavior coincide 
\cite{yos}, the use of the restricted operators is irrelevant as the occupation of the localized orbital is flat.

\section{Summary and discussion}

Within the framework of quantum simulation, we considered 
a well known effect for electrons under strong magnetic fields. 
A great amount of experimental data has been accumulated in solid state devices 
(Ref.~\cite{yos}) and the most appealing feature lies in the universality of the results. 
The plateaux of the Hall resistivity as a function of the (strong) magnetic field, 
obtained for very clean samples yet containing some disorder, signal the presence of peculiar states 
with well defined fractional filling factors. The values of $\rho_{yx}$ on the plateaux
depend only on universal constants ($\hbar$ and $e$). Up to now, most of these special 
states still require a challenging explanation not always complete. 
The translation of this physics to finite systems of atoms and its interpretation is not always easy. 
As a first attempt, our goal is to understand the mechanism in small samples and 
try to extrapolate the results to larger systems. 
Notwithstanding the fact that we can only tackle with small number of atoms, 
our results can be experimentally tested, given the ability of the new technologies to deal with very small samples 
and to engineer tunable impurities. 

Summarizing, we have obtained a plateau in the Hall resistivity $\rho_{yx}/\Omega$ following the 
expected line of search suggested by the known mechanism in the case of electrons and real fields 
in the fractional quantum Hall regime. The appearance  of a flat region on $\rho_{yx}$ reveals the essence 
of our goal, namely that there is a transfer of atoms from localized to non-localized atoms as $B^*$ increases. 
We proved that it is necessary to have part of the system localized around an impurity.
The transfer from localized to extended states allows the simultaneous variation of $B^*$ and $\rho_d$ 
producing a constant value of $\rho_{yx}\sim B^*/\rho_d$ along certain interval of $\Omega$. 
Particle interaction and the presence of impurities are crucial ingredients, 
at odds with the case of the integer quantum Hall effect. 
The improvement of the visualization of the plateau is achieved by forcing the complete exclusion 
of the localized part, using a restricted version of the operators for the current and the perturbation. 

So far we have used the expressions involved in the transport equation, 
valid for macroscopic systems with uniform density. Namely, the resistivity (or the conductivity) 
have no space dependence, and the known experimental behavior (see Ref.~\cite{wil}) 
is well captured by the linearity of $\rho_{yx}$ against $B^*$ (aside plateau structures or phase transitions 
of the GS). In contrast, our system is finite and the mean density decreases with $B^*$ 
producing a non-linear behavior of $\rho_{yx}$ with $B^*$, which is a finite size effect.

For systems with larger number of particles (several hundreds or thousands of atoms), the presence of impurities 
represents a small perturbation in the Hamiltonian. Therefore, for large systems with impurities randomly distributed, we do not expect that the plateau depends on the position of the impurities. 
This is not the case for systems with a small number of atoms, as in this case the presence of the impurities cannot be considered a small perturbation and the plateau depends on the position of the impurity (see Fig.~\ref{fig3}). 
Indeed, the ground state is strongly modified and so is $\rho_{yx}$, as shown in Fig.~\ref{fig4}, 
with the bump before the plateau in Fig.~\ref{fig4} being due to the presence of the phase transition 
taking place as the angular momentum jumps from $L=N(N-2)$ to $L=N(N-1)$. 
The change of angular momentum modifies the resistivity and produces a bump partially overlapped with the plateau. 

\section{Acknowledgments}

We appreciate interesting discussions with Jean Dalibard, Carlos Tejedor, Manel Bosch and Bruno Juli\'a-D\'iaz. 
A.T. acknowledges support from the Italian PRIN ``Fenomeni quantistici collettivi: dai sistemi fortemente correlati ai simulatori quantistici'' (PRIN 2010\_2010LLKJBX) 
and the Universitat de Barcelona for the kind hospitality. 
We acknowledge partial financial support from the DGI (Spain) Grant No. FIS2011-24154 and FIS2013-41757-P; 
the Generalitat de Catalunya Grant No. 2009SGR-1003, 2014SGR940 and 2009SGR21. We acknowledge also support from EU
grants OSYRIS (ERC-2013-AdG Grant No. 339106), SIQS
(FP7-ICT-2011-9 No. 600645), EU STREP QUIC (H2020-FETPROACT-
2014 No. 641122), EQuaM (FP7/2007-2013 Grant No.
323714), Spanish Ministry grant FOQUS (FIS2013-46768-P),  the Generalitat de Catalunya project 2014 SGR 874,
and Fundaci\'o Cellex.
J.T. is supported by grants FPA2013-46570, 2014-SGR-104 and Consolider
grant CSD2007-00042 (CPAN). Financial support from the Spanish Ministry of Economy and Competitiveness,
through the €˜Severo Ochoa€™ Programme for Centres of Excellence in R\&D (SEV-2015-0522) is acknowledged.

\appendix

\section{Linear response theory}
\label{sec:appendixLRT}

Following the standard protocol to analyze the linear response of a system~\cite{pit}, 
given an operator $\,-\lambda \hat{G} f(t)\,$ simulating a small perturbation (being $\hat{G}$ time-independent) 
and an operator $\hat{F}$ for a measurable quantity, the 
dynamical evolution of the expected value of the observable $\hat{F}$ is given by
\begin{equation}
\label{eq:j}
  \langle \hat{F}\rangle_t \,\,-\,\,\langle \hat{F} \rangle_0 \,=\, \lambda f(t) |\chi(\omega)|,
\end{equation}
where $\chi(\omega)$ has been defined as
\begin{eqnarray}
\label{eq:chi}
\chi(\omega) & = & \sum_{\nu\ne 0} [ \,\frac{\langle 0|\hat{F}|\nu\rangle \langle \nu|\hat{G}|0\rangle}{E_\nu-E_0+\omega+i\eta}
\nonumber
\\
& + & \frac{\langle 0|\hat{G}|\nu\rangle \langle \nu|\hat{F}|0\rangle}{E_\nu-E_0-\omega-i\eta}\,\,]\,\equiv\,|\chi(\omega)|e^{i\delta(\omega)}.
\end{eqnarray}
The sum is extended to all excitations and $\eta$ is a small quantity.

To obtain information about the Hall response and simulate Eq.~(\ref{eq:transport}), 
connecting with Section II, we make the following choice
\begin{equation}
\hat{G}=\sum_{i=1}^N \hat{x}_i,	
\nonumber
\end{equation}
\begin{equation}
\hat{F}=\hat{j}_y,	
\nonumber
\end{equation}
and
\begin{equation}
f(t)=\xi(t) sin(\omega t),	
\end{equation}
see Eqs.~(\ref{eq:pertH}),~(\ref{eq:xi}), and~(\ref{eq:jy}). If we identify the perturbation with 
the electric potential associated with a space independent electric field directed along the $X$-axis,
\begin{equation}
E_x(t)=\lambda f(t)
\end{equation}
then, from Eq.~(\ref{eq:j}) it is 
\begin{equation}
\sigma_{yx}=|\chi(\omega)|
\end{equation}
and from Eq.~(\ref{eq:resistivity}) we obtain the transverse resistivity $\rho_{yx}$.
The function $\chi(\omega)$ obtained numerically 
fulfills the condition $\chi^*(\omega)=\chi(-\omega)$ as well as the Kramers-Kronig relations~\cite{pit}.

\begin{figure}[tbp]
 \begin{tabular}{cc}
\includegraphics[width=0.47\columnwidth]{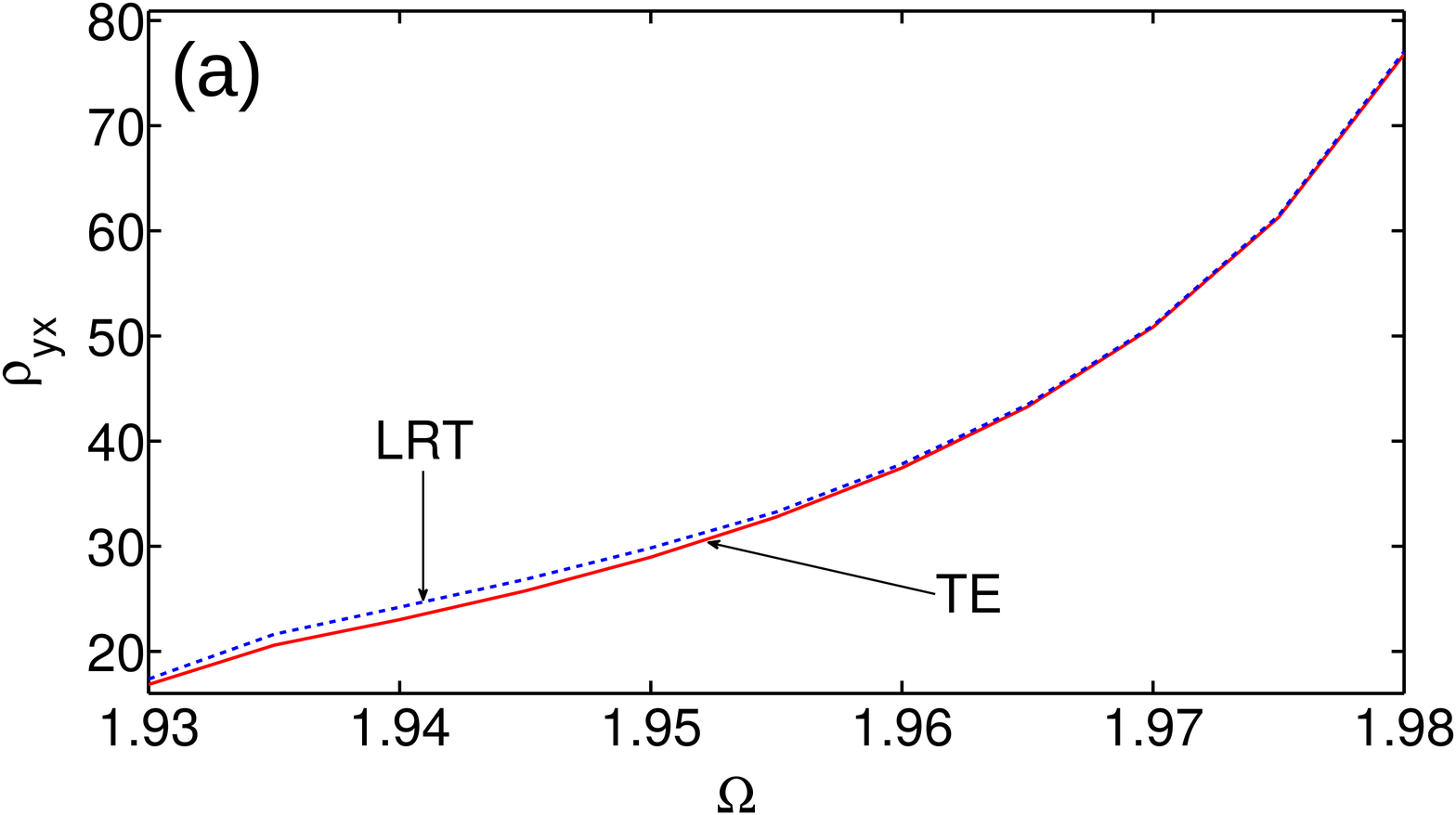}&\includegraphics[width=0.47\columnwidth]{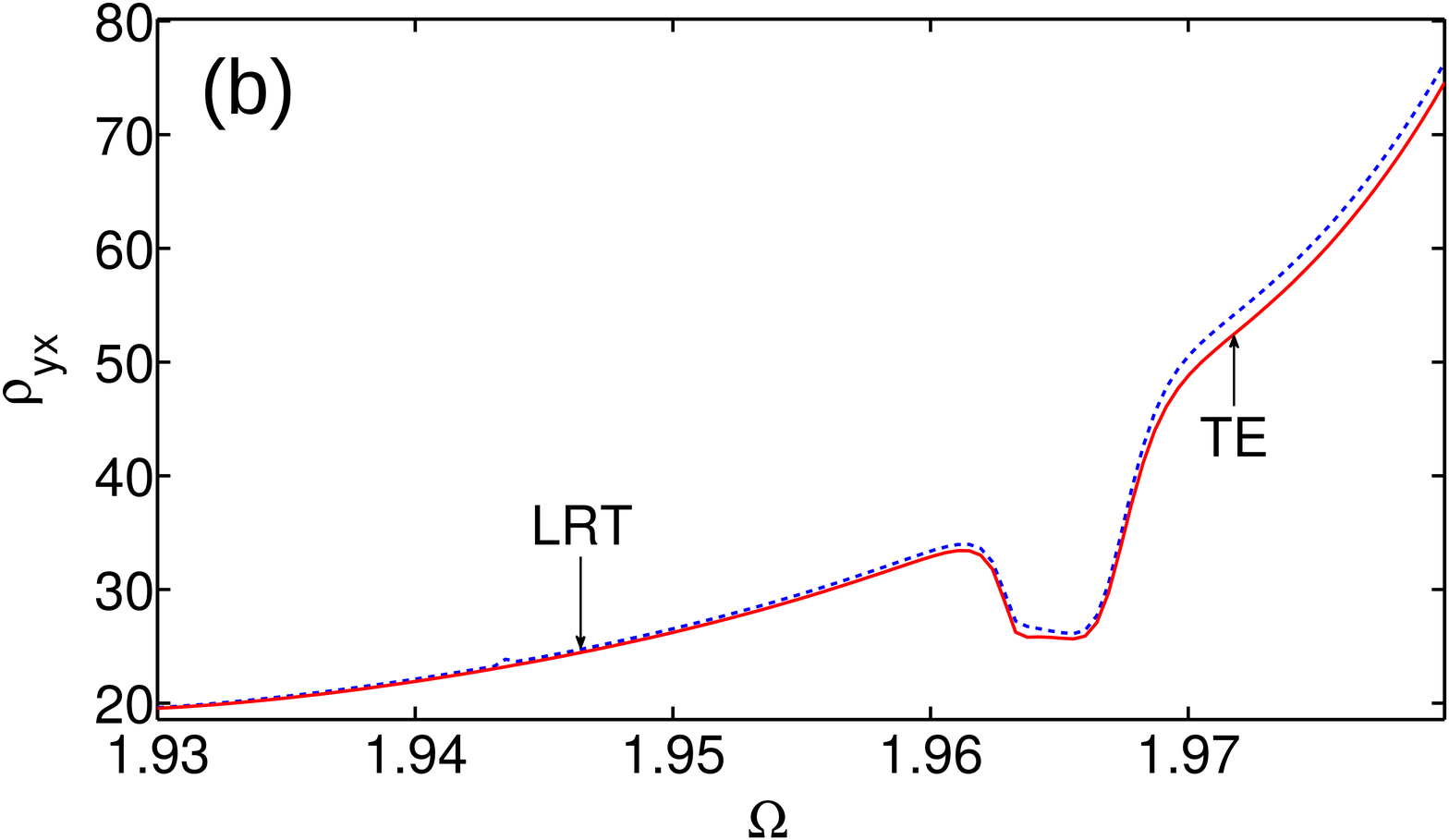}
\end{tabular}
\caption{ (a) Comparison between the calculation of the Hall resistivity $\rho_{yx}$ as a function of $\Omega$  with the complete dynamical evolution (TE) and the result from the LRT for $N=3$ and 
$\bf a_1=(1,0)$. We use here the restricted operators. Notice that for $N=3$ within this range of $\Omega$ no plateaux occur. (b) same fr $N=4$  when a plateau is apparent.  The figure shows that even in this case, we obtain nearly the same result.} \label{fig5}
\end{figure}

\begin{figure}[tbp]
\centering
\includegraphics[width=0.5\columnwidth]{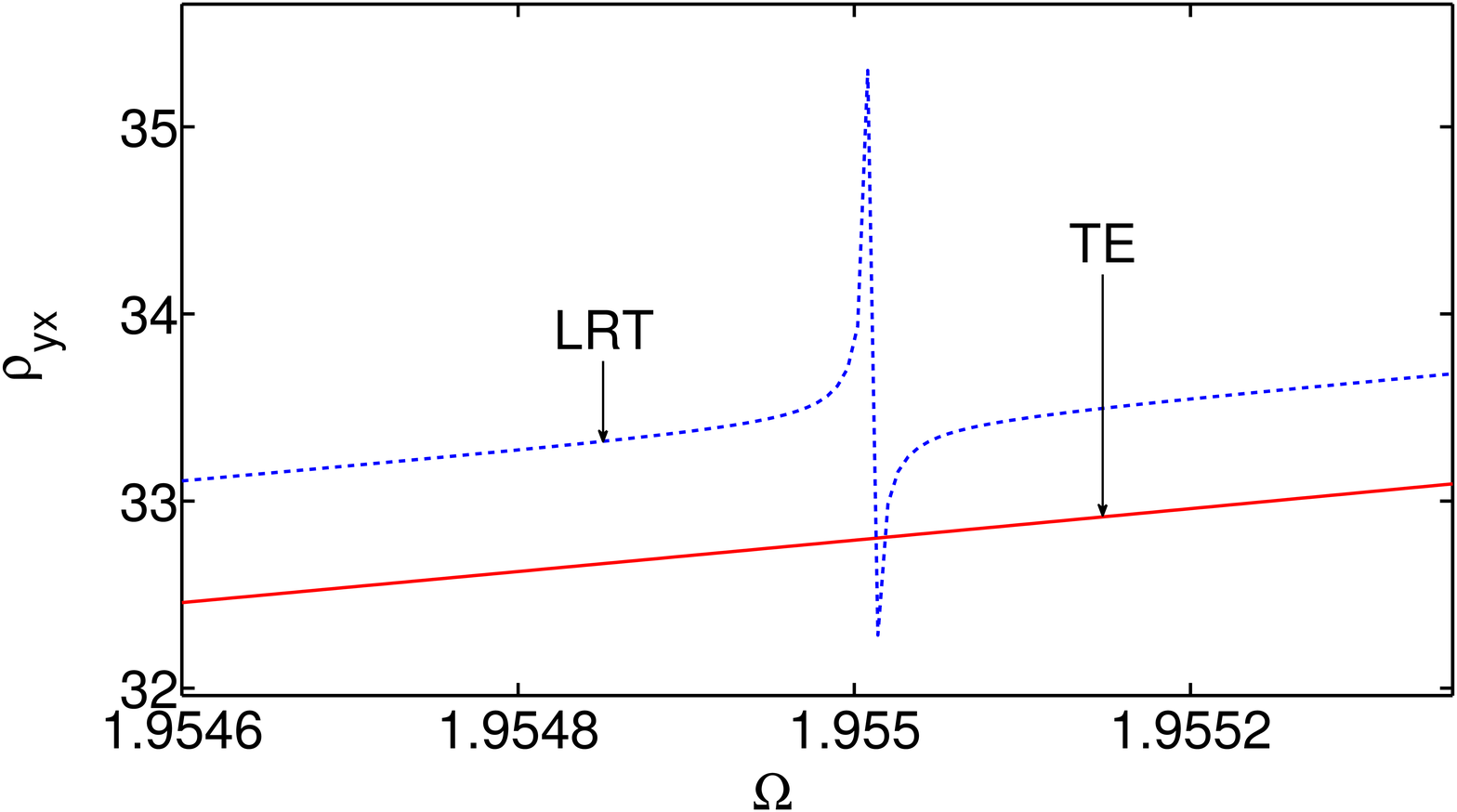}
\caption{  Hall resistivity $\rho_{yx}$ as a function of $\Omega$   for $N=3$ and 
$\bf a_1=(1,0)$ . This figure shows a typical resonant structure in the LRT calculation 
(upper curve) which disappears in the TE case (lower curve).}   \label{fig6}
\end{figure}

In what follows we consider three significant results. Fig.~\ref{fig5} (a) shows a comparison 
between the two techniques: TE and the LRT for $\rho_{yx}/\Omega$. 
This result indicates that, for a wide range of $\Omega$, the Hall effect is a linear response process. 
Figure~\ref{fig5} (b) is again a verification that the Hall response is linear 
even in the case where the response implies the formation of a plateau. 
Finally, we include Fig.~\ref{fig6} to illustrate possible limitations of the LRT. 
Notice that the structure of the upper part of Fig.~\ref{fig6} is the typical result of a resonance: 
For a critical value of $\Omega$ and for some excitations, a denominator 
in the expression of $\chi(\omega)$ [see Eq.~(\ref{eq:chi})] nearly vanishes. 
The system absorbs energy and the behavior is not linear. Namely the LRT is not applicable there. 
It could well be a sign of a peculiar state and its Hall response. 
However, the lower part of the figure demonstrates that this is not the case, 
that it is not related with the phenomena we are looking for. 
Notice that Fig.~\ref{fig5} (a) is for $N=3$, in this case, $\rho_{yx}/B^*$ does not present a plateau, 
at odds with the result shown in Fig.~\ref{fig5} (b) for $N=4$ where within the same range of $\Omega$ values, 
a plateau is apparent. In Fig.~\ref{fig5} the LRT result does not show the resonance due to the low resolution.

\end{document}